\documentclass[12pt,a4paper,twoside,english]{article}
\usepackage[T1]{fontenc}
\usepackage[latin1]{inputenc}
\usepackage{amssymb}

\makeatletter

\usepackage{babel}
\makeatother

\begin{document}

\title{A new approach to the quantized electrical conductance}

\author{{\normalsize M. Apostol }\\
{\normalsize Department of Theoretical Physics, Institute of Atomic
Physics, }\\
{\normalsize Magurele-Bucharest MG-6, POBox MG-35, Romania }\\
{\normalsize email: apoma@theory.nipne.ro}}

\date{{}}

\maketitle
\relax

\begin{abstract}
The quanta of electrical conductance is derived for a one-dimensional
electron gas both by making use of the quasi-classical motion of a
quantum fluid and by using arguments related to the uncertainty principle.
The result is extended to a nanowire of finite cross-section area
and to electrons in magnetic field, and the quantization of the electrical
conductance is shown. An additional application is made to the two-dimensional
electron gas.
\end{abstract}
\relax



\noindent Recently, there is a considerable deal of interest in the
quantized electrical conductance of atomic and molecular conductors
like nanowires, narrow atomic constrictions, quantum dots, carbon
nanotubes, etc.\cite{key-1}-\cite{key-11} The effect was originally
predicted by Landauer.\cite{key-12}-\cite{key-14} We present here
a new derivation of the quanta of electrical conductance for a one-dimensional
electron gas, by making use of two procedures: the quasi-classical
approach to the one-dimensional quantum electron fluid and by using
arguments related to the uncertainty principle. We extend the results
to the quantization of the electrical conductance in a quasi one-dimensional
nanowire of finite area of the cross-section, where the electron motion
is confined to the transversal directions while the free longitudinal
motion is subjected to the action of the electric field. An application
of this result is made for the two-dimensional electron gas. Also,
we apply the present approach to electrons in a magnetic field. 

We consider first a one-dimensional (free) electron gas in a conductor
of length $l$ and cross-section area $A$. We consider a purely quantum
transport in such a conductor, without scattering or thermal effects.
The electron density is given by $n=gk_{F}/\pi A$, where $k_{F}$
is the Fermi wavevector and $g$ is a degeneracy factor (\emph{e.g.}
$g=2$ for spin $1/2$). In the presence of an electric field $E$
along the conductor the density is modified at the Fermi level by
$\delta n$, such that, locally, higher energy levels are occupied
for the electrons moving oppositely the field and Fermi energy levels
are depleted for electrons moving along the field. We take the field
oriented along the negative $x$-direction, so the net flow of electrons
takes place along the positive $x$-direction. The electric field
is sufficiently weak and slowly varying such that the electrons acquire
a displacement $u(x)$ which obeys the quasi-classical equation of
motion $m\ddot{u}=eE$, where $m$ is the electron mass and $-e$
is the electron charge. The change in the electron density is given
by $\delta n=-n\partial u/\partial x$, such that the density of electrons
participating in the electrical flow is $-\delta n$. From these two
equations we get straightforwardly \begin{equation}
m\frac{d}{dt}\delta n=-enE/v_{F}\,\,\,\label{1}\end{equation}
 for a constant field, where $v_{F}$ is the Fermi velocity. This
is the basic equation for computing the electrical current.%
\footnote{The quasi-classsical motion of the one-dimensional quantum electron
gas was previously discussed in more detail in Ref. 15. %
} Indeed, the electrical flow (charge per unit area of the cross-section
and per unit time) is given by $j=-e(-\dot{\delta n})l=-e^{2}nEl/mv_{F}$,
hence the well-known electrical conductivity $\sigma=e^{2}nl/mv_{F}$.
The electrical flow is negative, \emph{i.e.} it is oriented along
the electrical field as it should be. For the one-dimensional gas
$n=gk_{F}/\pi A$ and $v_{F}=\hbar k_{F}/m$, so we get $\sigma=g(2e^{2}/h)(l/A)$,
where $h$ is Planck's constant ($\hbar=h/2\pi$). The electrical
conductance is $G=\sigma A/l=g(2e^{2}/h)$. It can be written as $G=\sum_{s}G_{0}$,
where $s$ is the spin variable (\emph{e.g.} $s=\pm1$ for spin $1/2$)
and $G_{0}=2e^{2}/h$. We can see that the electrical conductance
can only vary by quanta $G_{0}$, acccording to spin degeneracy. 

It is worth noting that the same result can be obtained by applying
the uncertainty principle. Indeed, the equation of motion $md\dot{u}/dt=eE$
can also be written as $m\dot{u}=eE\tau$, where $\tau$ is the time
of motion. The electrical flow can be written as $j=-e\dot{u}\delta n$,
where $\delta n=g\delta k_{F}/\pi A$ is the density of electrons
participating in conduction. Combining these two equations we get
$j=-e^{2}E\tau\delta n/m$ and $\sigma=e^{2}\tau\delta n/m=ge^{2}\tau\delta k_{F}/\pi Am$,
which is another representation for the electrical conductivity. Now
we use the uncertainty principle in the form $\tau=\delta n_{F}(h/\delta\mathcal{E})$,
where $\delta n_{F}=l\delta k_{F}/2\pi$ and the change in energy
is given by \begin{equation}
\delta\mathcal{E}=\frac{\hbar^{2}}{2m}(k_{F}+\delta k_{F})^{2}+\frac{\hbar^{2}}{2m}(k_{F}-\delta k_{F})^{2}-\frac{\hbar^{2}}{m}k_{F}^{2}=\frac{\hbar^{2}}{m}(\delta k_{F})^{2}\,\,.\label{2}\end{equation}
The motion time given by the uncertainty principle corresponds to
$\delta n_{F}$ cycles of quanta of action $h$. The change in energy
given by (\ref{2}) is also $\delta\mathcal{E}=-eV$, where $V$ is
the voltage drop, which shows that the voltage is also quantized.
We consider the energy levels sufficiently dense as to allow a continuous
change in the electrical potential. Combining all these formula given
above we arrive again at the conductance $G=gG_{0}$. 

Within the quasi-classical description by means of the displacement
field $u$ the electrical field is given by $E=d\varphi/du$, where
$\varphi$ is the electrical potential, such that the equation of
motion $m\ddot{u}=eE$ ensures the conservation of energy. This equation
of motion can also be written as $\dot{\Pi}=eE=ed\varphi/du$, or
$\dot{u}=ed\varphi/d\Pi$, where $\Pi=m\dot{u}$ is the momentum associated
to the field $u$. One may check indeed that $\Pi=\hbar\delta k_{F}$
by making use of equation (\ref{2}), so we may also write $\dot{u}=ed\varphi/dp_{F}=ed\varphi/\hbar dk_{F}$.
Now the electrical flow $j=-e\dot{u}\delta n$ becomes $j=-e^{2}(d\varphi/\hbar dk_{F})\delta n=-e^{2}(dn/\hbar dk_{F})V$
(since $\delta n=(dn/d\varphi)V$), hence the electrical current $I=jA=-gG_{0}V$.
This was, in essence, the original argument of Landauer.\cite{key-14}
One can see that the conductance is proportional to the density of
states.%
\footnote{See in this respect Ref. 3. %
}

We consider next a nanowire of thickness $d$ ($A=d^{2}$) and a confined
transversal motion of the electrons, such that the energy levels are
given by \begin{equation}
\varepsilon=\frac{\hbar^{2}k^{2}}{2m}+\frac{\pi^{2}\hbar^{2}}{2md^{2}}(n_{1}^{2}+n_{2}^{2})\,\,\,,\label{3}\end{equation}
 where $n_{1,2}$ are positive integers.%
\footnote{We impose fixed-ends boundary conditions for the transversal motion
as for electrons confined in an infinite potential well (see Refs.
1-3).%
} We have now multiple branches of one-dimensional electron gas and
the Fermi wavevector depends on the duplex $(n_{1},n_{2})$. Therefore,
the electron density is given by \begin{equation}
n=(g/\pi A)\sum_{(n_{1},n_{2})}k_{F}(n_{1},n_{2})\,\,\,\label{4}\end{equation}
and the electrical conductivity $\sigma=e^{2}nl/mv_{F}$ becomes \begin{equation}
\sigma=(e^{2}l/\pi Am)\sum_{(n_{1},n_{2}),s}k_{F}(n_{1},n_{2})/v_{F}(n_{1},n_{2})\,\,.\label{5}\end{equation}
By (\ref{3}), the Fermi velocity is $v_{F}=\hbar k_{F}/m$, such
that the above electrical conductivity becomes $\sigma=G_{0}(l/A)M$
and the electrical conductance is quantized according to $G=G_{0}M$,
where \begin{equation}
M=\sum_{(n_{1},n_{2}),s}1\,\,\,\label{6}\end{equation}
 is the number of branches in the electron spectrum (number of channels),
spin included. One can see that $M$ is the number of channel electron
states, so the conductance can only change in steps of quanta $G_{0}$.

Now we want to compute the number of channels $M$ for this model.
We assume a dense distribution of spectrum branches and write $n_{1}^{2}+n_{2}^{2}=\rho^{2}$
in equation (\ref{3}). The chemical potential $\mu$ is established
by the equalities \begin{equation}
\mu=\frac{\hbar^{2}k_{F}^{2}}{2m}+\frac{\pi^{2}\hbar^{2}}{2md^{2}}\rho^{2}\,\,\,,\label{7}\end{equation}
hence the Fermi wavevector $k_{F}$ which is used in equation (\ref{4}).
The number of channels is then given by $M=\pi gN_{t}^{2}$, where
$N_{t}$ is the highest integer $\rho$ satisfying equation (\ref{7}).
It is given approximately by $N_{t}^{2}=2md^{2}\mu/\pi^{2}\hbar^{2}$.
Equation (\ref{4}) can then be written as \begin{equation}
n=(2g/A)\int_{0}^{N_{t}}d\rho\cdot\rho\sqrt{2m\mu/\hbar^{2}-\pi^{2}\rho^{2}/d^{2}}\,\,.\label{8}\end{equation}
This equation gives a relationship between $N_{t}$ and $\mu$, which,
together with the equation $N_{t}^{2}=2md^{2}\mu/\pi^{2}\hbar^{2}$
written above, serve to determine both the chemical potential $\mu$
and the number $N_{t}$ of transverse channels, hence the total number
of channels $M$, as a function of the density of the electron gas.
The integral in equation (\ref{8}) can be performed straightforwardly.
We get $N_{t}=(3Nd/\pi gl)^{1/3}$ and $M=(\pi g)^{1/3}(3Nd/l)^{2/3}$,
where $N$ is the total number of electrons and $l$ is the length
of the sample. The electrical conductance reads $G=G_{0}(\pi g)^{1/3}(3Nd/l)^{2/3}$.
This is the main result for the overall continuous behaviour of the
quantized conductance of an ideal nanowire of finite cross-sectional
area.

A simple application of the above result pertains to a two-dimensional
electron gas with point contacts.\cite{key-10} In this case we may
consider that the energy spectrum given by equation (\ref{3}) contains
only one quantum number, say $n_{1}$, as corresponding to the quantized
motion along one tranverse direction. The number of channels is now
$M=gN_{t}$, where $N_{t}$ is given by $\mu=\pi^{2}\hbar^{2}N_{t}^{2}/2md^{2}$.
The chemical potential $\mu$ can be obtained straightforwardly from
equation (\ref{4}) as $\mu=2\pi\hbar^{2}An/gmd$, which leads to
$M=gN_{t}=2\sqrt{gNd/\pi l}$. If we take, as ususally, $N=gW^{2}k_{F}^{2}/4\pi$,
where $W$ is the width of the contacts, we get $M=gk_{F}W(d/l)^{1/2}/\pi$,
which is a slight generalization of the classical result verified
experimentally in the quantized conductance $G=G_{0}M$ of a $GaAs-AlGaAs$
heterostructure with ballistic point contacts.\ref{10} 

As it is well-known, for electrons in a magnetic field $H$ we can
write the energy levels as \begin{equation}
\varepsilon=\frac{\hbar^{2}k^{2}}{2m}+\hbar\omega_{c}(n+1/2)+\mu_{B}Hs\,\,\,,\label{9}\end{equation}
where $\omega_{c}=eH/mc$ is the cyclotron frequency, $\mu_{B}=e\hbar/2mc$
is the Bohr magneton and $s=\pm1$. Equation (\ref{5}) gives now
$G=G_{0}M$, where \begin{equation}
M=\sum_{ns}1\,\,.\label{10}\end{equation}
It is approximately $M=2n_{l}$, which is in fact twice the number
of spectrum branches. We have also \emph{$\mu=\hbar\omega_{c}n_{l}$},
where  $\mu$ is the chemical potential given by \begin{equation}
n=\frac{4eH}{\pi ch}\int_{0}^{n_{l}}dn\sqrt{2m\mu/\hbar^{2}-2m\omega_{c}n/\hbar}\,\,\,,\label{11}\end{equation}
which is similar to equation (\ref{8}). Here it is worth noting the
well-known transversal degeneracy $2eHA/ch$ ($\gg1$) of the energy
levels in the magnetic field. We get $M=(3\sqrt{\pi}/4\sqrt{2})^{2/3}n^{2/3}ch/eH$
(which should be much larger than unity). This main result allows
the calculation of the quantized conductance $G=G_{0}M$. It is worth
emphasizing that the electrical conductance (or magnetoresistance)
can be varied in quantum steps by varying the magnetic field, as it
is well-known. 

The inclusion in such a treatment of interaction, scattering or thermal
effects (or finite-size boundary effects), as well as other particularities,%
\footnote{\emph{e.g.} for specular reflection in a cylindrical conductor in
magnetic field see Ref. 8. %
} renders the problem a bit more complicated. Generally speaking, the
starting point in such a treatment is the notion of elementary excitations
and their lifetime. Particularly interesting is this problem for multi-wall
carbon nanotubes, due to their specific electron energy structure.\cite{key-6}

\end{document}